\documentstyle[amssymb,12pt]{article}

\newcommand{\be}{\begin{equation}}
\newcommand{\ee}{\end{equation}}

\title{Soft Binary Processes, NJL Model and Absolute\\
Values
of the Amplitudes of Reactions\\ $\pi^-p \to \pi^0 n$
and $\pi^-p \to \eta n$}
\author{I.I.Levintov\\
State Scientific Center\\
Institute of Theoretical and Experimental Physics\\
117259, Moscow, Russia}
\date{}

\begin{document}
\maketitle

%\def\la{\mathrel{\mathpalette\fun <}}
%\def\ga{\mathrel{\mathpalette\fun >}}
%\def\fun#1#2{\lower3.6pt\vbox{\baselineskip0pt\lineskip.9pt
%\ialign{$\mathsurround=0pt#1\hfil##\hfil$\crcr#2\crcr\sim\crcr}}}

%\large
\section*{Abstract}
Reactions $\pi^-p \to \pi^0 n(\eta n); \pi^-p \to K^0 \Lambda(\Sigma)$ occur
through two phases: 1.~Gribov's diffusion of constituent quarks of each of
interacting hadrons in the space of rapidities and impact parameters with
production of a flux-tube which has a fast spectator on one edge and a slow
reagent on the other. This phase determines a power decrease of the
amplitude with energy increase.~ 2.~The charge-exchange of slow reagents
$\bar{u} u \to \bar{d} d;~ \bar{u} u \to s \bar{s}$ determines the value of
the residue of the Regge pole. Reaction $\pi^-p \to \pi^0 n(\eta n)$
contains: the scalar bilinear $(\bar{u} u)(\bar{d} d) = (^3P_0)^2$
determining the dominant spin-orbital amplitude  $M_1(^3P_0)$ and the
pseudoscalar bilinear $(\bar{u} i \gamma_5 u)(\bar{d} i \gamma_5 d) =
(^1S_0)^2$. In the amplitude $M_0~~ (^3P_0)^2$ and $(^1S_0)^2$ interfere
destructively and strongly. These facts which follow from the analysis of
experimental data agree with the NJL model predictions. Spontaneous chiral
symmetry violation leads to that the bilinear weights are independent of
the coupling constant of NJL Lagrangian (the "blackness" condition) and
are determined by the transversal dimension $R^2$ in the charge-exchange
region $\bar{u} u \to \bar{d} d$ entering the physical amplitude expression
as a radius of the residue of the Regge pole. $M_1(^3P_0)$ at small 
$q^2_{\bot}$ contains:  the colour number $N_c = 3;~ R^2$ and $\alpha(0)$. 
The comparison with the experimental $M_1(exper.)$ gives $\vert M_1(^3P_0) 
\vert / \vert M_1(exper) \vert = 1.09(1\pm 0.1);~ 0.7(1 \pm 0.2)$ for 
reaction $\pi^- p \to \pi^0n; \eta n$, correspondingly within the interval 
$s = 8-400 GeV^2$ and $0.004 \leq q^2_{\bot} \leq 0.1 (GeV/c)^2$.

In the framework of the formally $SU(3)$ symmetrical NJL model we discuss
the reason of a strong SU(3) symmetry violation which manifests itself in the
observable smallness of $M_1$ in reactions $\pi^-p \to K^0 \Lambda(\Sigma)$.

\section{Introduction}

I will consider the charge-exchange reactions $0^- \frac{1}{2} \to
0^-\frac{1}{2}~~~ \pi^- p \to \pi^0 n (1), ~~ \pi^-p \to \eta n
(2), ~~ \pi^-p \rightarrow K^0 \Lambda (\Sigma) (3)$ in the Regge region of
$s$ and $t$. Figs.1 a,b and 2 show some data on differential cross sections
of these reactions [1,2,3]. The amplitudes $M$ and differential cross
sections are expressed via elements $M_0$ (non-flip) and $M_1$ (spin-flip):
%1
\be
M = M_0 + i \vec{\sigma} \vec{n}  M_1 \frac{q_{\bot} R}{2};~~ 64 \pi p^2_c s
\frac{d \sigma}{d q^2_{\bot}} = \vert M_0 \vert^2 + \vert M_1 \vert^2
\frac{q^2_{\bot} R^2}{4}
\ee
where $\vec{\sigma}$ are Pauli matrices,
$\vec{n}$ is an orth of normal of the reaction plane, $q_{\bot}$ is
transverse momentum transfer, $p_c$ is momentum in the c.m.s, $R$ is the
radius of the residue of Regge poles which dominates in the reactions.

Our choice is due to simple spin structure of these reactions and to
presence of exact data on differential cross sections of reactions 1,2
within the interval $s = 8 - 400 GeV^2;~~ q^2_{\bot} = 0.004 -
0.3(\frac{GeV}{c})^2$ [1,2].

The simplest scheme of reactions 1,2,3 is presented by the
dual diagram of Fig.3b, which is interpreted by A.B.Kaidalov in [4] as a
three-stage process [5] of Fig.3c. In stage I choose configurations of
initial hadrons with small rapidities of annihilating constituents. These
configurations fuse into intermediate flux-tube (f.t) where the transition
$\bar{u} u \rightarrow \bar{d}d$ or $\bar{u} u \to s \bar{s}$ (stage II)
appear, after that f.t. disintegrates into two parts which  evolve into
finite hadrons (stage III) or continue disintegrating giving rise to a
multiple process.

This picture is based on experimental facts:

1. Power decrease of cross-sections with $s$ increasing
\be
\frac{d \sigma}{d t} ch.ex. \sim s^{2[\alpha_R(-t) -1]};~~ \alpha_R(-t) < 1
\ee
where $\alpha_R$ -- is trajectory of the secondary Regge pole. The power
decrease of cross sections agrees with the parton concept providing the
rapidities of annihilating constituents to be small and independent of the
rapidity of initial hadrons [4,6].

2. Mass relations [4] which follow from factorization of the binary amplitude
in $s$ channel are well fulfilled. Factorization in $s$ channel is a good
argument in the favour of existence of an intermediate object -- f.t.

3. The appearance of f.t. with increase of colour dipole moment of a pair
$\bar{q} q$ static test quarks is confirmed in QCD on lattices [7].

The model [4] results in a correct Regge asymptotics of binary processes
[8].

In the framework of the dual diagram of Fig.3b the spin structure and the
value of residues of physical amplitude poles is determined by the spin
structure of soft 4-quark interaction in a subprocess accompanied by a
change of flavour $\bar{u} u \rightarrow \bar{d} d,~~ \bar{u} u \rightarrow
s \bar{s}$ which holds in f.t. in the II stage and may be represented by
five 4-fermionic bilinear amplitudes [6]
%3
\be
(2 p_e \sqrt{s})^{-1} Im M_b = \langle f.t  \vert \sum \limits^{5}_{n=1} ~
T_n(\bar{q}_i q_i \rightarrow \bar{q}_f q_f) \vert f.t. \rangle
\ee
where symbol $\langle f.t. \vert T \vert f.t. \rangle$ implies the soft pair
charge-exchange amplitude averaged over relative moment of $\bar{q}$ and
$q$. The pair results (by analogy with [4]) from diffusion of partons of
interacting hadrons in the space of rapidities and impact parameters
$\vec{b}$ [9]. The weights and signs of bilinears $T_n$ are given by the
effective chiral Nambu and Jona-Lasinio Lagrangian (the NJL model [10]),
broadened by 't Hooft [13] (see
reviews [11], [12].

In this paper I will show that such a construction well reproduces the
experimental data of reactions 1 and 2.

Since in the NJL model, in the lowest in $1/N_c$ order ($N_c$ is the colour
number) there are only bilinears of scalars ($s$) and pseudoscalars ($p.s.$)
[12]:
%4
$$
(2 p_c \sqrt{s})^{-1}~ Im M_b = \langle f.t. \vert T (S) - T(P.S) \vert f.t.
\rangle =
$$
\be
\langle f.t. \vert (\bar{q}_i q_i) D_s (\bar{q}_f q_f) - (\bar{q}_i i
\gamma_5 q_i)~ D_{ps} (\bar{q}_f i \gamma_5 q_f) \vert f.t. \rangle
\ee
where $D$ are propagators of intermediate states. The sign $-$ in eq.(4) is
determined by the sign in the 't Hooft determinant not conserving SU(3)
singlet current $(\bar{u} u + \bar{d} d + \bar{s} s)$.

The account of bilinear $(\bar{u} u)(\bar{d} d)$ in eqs.(3), (4) makes it
possible to explain qualitatively a large spin-orbital effect in reactions
1,2 , their planarity and smallness of spin-orbital interaction at
the Pomeranchuk pole [6, 14]. Let us remind this explanation.

Fig.3d shows the $^3P_0 = O^+$ scheme of reaction $\pi^-p \rightarrow
\pi^0n$ on a "proton" consisting of a totally polarized along $+z$ axis soft
$u$ quark and a leading spectator $u d$ with the impact parameter $+b_{ud}$.
The impact parameter of mesonic spectator -- $b_d = 0$. The sign 
$\curvearrowleft$
denotes the transversal spin of quark with $s_z = +\frac{1}{2}$. Z axis is
directed along the normal of reaction plane. In $^3P_0$ state the
transverse orbital moment $l_z = -1$ denoted as 
$\rightleftarrows$ 
and
the summary transversal spin $\bar{u} u~~ S_z = +1$ are antiparallel. That
is why the quark charge-exchange, under a nontriviall condition
of $S_z$ and $l_z$ conserving in $^3P_0 = 0^+$ state in the $\bar{u} u
\rightarrow \bar{d} d$ process follows predominantly at $+b$, since at $-b$
annihilation in $^3P_0$ state is realized with a smaller probability ($l_z$
and $S_z$ are parallel). Thus, $M(+b) - M(-b) = M_{l -\frac{1}{2}} - M_{l
+\frac{1}{2}} \not= 0$, i.e. $\vec{l} \vec{\sigma}$ interaction appears. In
the third part of the paper I show that  it is $T(S) = T(^3P_0)$ but not
$T(^3P_2)$ which determines $M_1$ in reaction 1. At P pole effects due to
soft quark annihilation are of no importance and spin-orbital interaction is
small.

Large spin-orbital amplitude $M_1$ in reaction 1 is a result of $l_z$ and
$S_z$ conservation  in  $^3P_0 = 0^+$ state in the $\bar{u} u \to \bar{u}
d$ process. We deal with a "hidden orbital moment". Conservation of $l_z$
corresponds to the planary process $\bar{u} u \to \bar{d} d$ (Fig.4). Thus
the presence of a dip in reaction is an experimental evidence for planarity
and, consequently, for that dual diagram is appropriate to describe $\pi^-p
\to \pi^0n$. Some arguments according to which all the mentioned on reaction
1 is also valid for reaction 2, are presented in the third part of the
paper.

The next discussion is planned as follows:

In the 2-d part we consider some consequences of extended NJL model for the
$\bar{u} u \to \bar{d} d$ amplitude. One may expect them to reveal in the
$M_{0,1}$ amplitudes in reactions 1 and 2. We note

1. Dominance of $S$ and $PS$ bilinears in (3).

2. Their strong destructive interference in $M_0$ provided that the
intermediate particle mass in P.S. channel is small.

3. Independence of $\bar{u} u \to \bar{d} d$ soft amplitudes on the coupling
constant of the NJL effective Lagrangian stemming from spontaneous violation
of chiral symmetry. This fact is an argument in the favour of existence of
the "blackness condition" (12), the physical meaning of which is that the
soft $\bar{u} u \to \bar{d} d$ amplitude is determined by transversal
dimension of $R$ region where the pair charge-exchange occurs, but not by the
coupling constant. In the 4-th section we see that $R$ enters the expression
for $M$ as a radius of residue of the Regge pole.

In the 3-d section we perform the analysis of experimental data based on
the Fiertz expansion of the $t$ channel 4-quark amplitude in the dual
diagrams of reactions 1 and 2 taking into account their exchange
degeneration. This analysis will confirm the 1-st and the 2-d consequence
of the NJL model for reactions 1 and 2 (strong destructive interference of
$M_0(^3P_0)$ and $M_0(^1S_0)$ and, consequently, the smallness of the
intermediate particle mass in P.S.channel and shows that in the framework of
dual diagram $M_1 = M_1(^3P_0)$. The experimental
$\vert M_1 \vert /\vert M_0 \vert = 5; 10$ for reactions 1 and 2
,respectively,are compared with the estimate ($\vert M_1 \vert
/\vert M_0 \vert)_s < 1)$ in reactions $\pi^-p \to K^0 \Lambda (\Sigma)$
obtained on the basis of the differential cross sections data.It is stated
that the dynamics of $\bar{u} u \to \bar{d} d$ and  $\bar{u} u \to s
\bar{s}$ is different. There is a strong SU(3) symmetry violation in $M_1$.
Possible reasons for this phenomenon are discussed within the NJL model.

In the 4-th section the relation (3) is written in explicit form  (29)
basing on the unitarity condition of the f.t. model in $s$ channel. With the
account of the "blackness" condition (12) this enables one to calculate
absolute values of the $M_1(^3P_0)$ amplitudes which are dominant in
reactions 1 and 2. The model expression $M_1(^3P_0)$ (39) contains: the
colour number $N_c = 3$, the radius of the residue of the Regge pole $R^2$
and $\alpha(-t)$. Since $R^2$ and $\alpha(-t)$ are determined from the
energy dependence of the slope of reactions cone, the expressions for the
absolute value $M_1(^3P_0)$ has no free parameters $M_1(^3P_0)$ is compared
with the experim ental values $M_1$(exper.) (24). As a result:

$$\vert M_1(^3P_0) \vert / \vert M_1(\mbox{exper.}) \vert = 1.09 (1 \pm 0.1);
~~ 0.7(1 \pm 0.2)$$

for reactions 1 and 2, correspondingly. This result describes the
experimental data within the intervals: $0.004 \Biggl (\frac{GeV}{c}
\Biggr )^2 \leq q^2_{\bot} < 0.1 \Biggl (\frac{GeV}{c} \Biggr )^2$ and $s =
8 - 400 GeV^2$.

We have refrained from estimating the $M_0$ amplitudes because of their
theoretical ambiguity: $M_0 = M_0(^3P_0 - M_0(^1S_0)$ is a small difference
of large values. The 5-th concluding part deals with some problems following
from the results of our analysis.

\section{Extended NJL Model and Soft $\bar{u} u \to \bar{d} d$ and
$\bar{u} u \to s \bar{s}$ Amplitudes.}
The minimal $U_L(1) \otimes U_R(1)$ symmetric (chiral) NJL Lagrangian of the
$\bar{q} q$ light pair system with one flavour in Dirac vacuum has
the form
%5
\be
{\cal L}_{NJL} = \bar{q} (i \gamma \partial - m) q + G \Biggl 
[ (\bar{q} q)^2 + (\bar{q} i \gamma_5 q)^2 \Biggr ]
\ee
where $G$ is the positive interaction constant (quarks undergo attraction),
$m = 5 MeV$ - is the current quark mass, $\gamma$ is the Dirac matrix, $q$
is the quark field operator. The form of this bare Lagrangian is determined
by requirement of chiral invariance. The current quark mass is small as
compared with the cut-off energy of nonrenormalizable NJL model $\Lambda
\simeq 1 GeV$. The value $\Lambda = 0.631 GeV$ [12] is agreed with the
low-energy data array.

The equality of interaction constants with which bilinears of scalar and
pseudoscalar currents enter ${\cal L}_{NJL}$ is the fact following from requirement
of chiral invariance. This equality takes place also for extended NJL model.

When $G$ increases up to a critical value $G_c \geq \pi^2/N_c \Lambda^2$
the current quark system becomes energy-unstable. The chiral symmetry
spontaneously violates as a result of the instanton transition and phase
transition into minimal energy state occurs. Here current quarks of each
flavour acquire mass (in the extended NJL model due to scalar virtual
$\bar{q} q$ pairs of three flavours) and go over to dynamical (constituent)
quarks with the mass $M_c \simeq 0.3 GeV$ which are paired into scalar pairs
with $J^{\pi} = 0^+$ producing a new "QCD" vacuum of chiral condensates
$<\bar{q} q>$ with the energy density $\epsilon < 0$.

An energetic gap $\vert \Delta E \vert \geq \vert \epsilon \vert + M_c$
appears between external constituent quarks and the new vacuum. Since the
NJL model considers only fermionic degrees of freedom chiral condensate is a
condensate of constituent $^3P_0$ pairs (by analogy with super-conductor).

Production of massive $^3P_0$ pairs in which transversal spin of a $\bar{q}
q$ is compensated by transversal orbital moment, from light pairs of
longitudinal current quarks with opposite moments and equal chiralities
(bilinear scalar in (5)) is connected with quark spin flip described by the
Bogolyubov-Valatin transformation and should be accompanied by production
of pseudoscalar Goldstone bosons  (bilinear pseudoscalar in (5)) which are
associated with an octet of pseudoscalar mesons.

On the background of new vacuum (chiral condensate) external constituent
quarks undergo residual interaction which may be represented, under mean
field approximation, by the sum of two terms of the extended NJL model:
%6
$$
{\cal L}^{res}_{NJL} = {\cal L}^{res}_{sym} + {\cal L}^{res}_{det}
$$
\be
{\cal L}^{res}_{det} = G_{t H} \Biggl [det~ \bar{q}_i(1 + \gamma_s) q_f +
det~\bar{q}_i (1 - \gamma_s) q_f \Biggr ]
\ee
The first term has the structure of (5), it does not mix flavours, if
small effects due to current mass difference will be excluded. The
determinant term constructed by 't Hooft violates $U_A(1)$ symmetry, does
not conserve singlet $SU_3$ current and consequently gives transitions
$\bar{u} u \to \bar{d} d, \bar{u} u \to s \bar{s}$. In the case of three
flavours ${\cal L}^{res}_{det}$ has six-fermion structure.
${\cal L}^{res}_{det}$ can be reduced to the effective 4-quark
representation [15] which imitates instantonous transition
(Fig.5).
%7
\be {\cal L}^{res}_{t H} (4) = <\bar{q}q > K \sum
\limits^{9}_{n=0} \Biggl [(\bar{q}_i \lambda_n q_i) 
(\bar{q}_f \lambda_n q_i) - (\bar{q}_i i
\gamma_5 \lambda_n q_i) (\bar{q}_f i \gamma_5 \lambda_n q_f) \Biggr ]
\ee
where
$\lambda_0 = - \sqrt {2/3} \cdot I, ~~~ \lambda_n (n = 1-8)$ are Gell-Mann
matrices [16], (7) gives the sign $-$ in (4).

For the case of two flavours ($u$ and $d$) and up to $1/N^2_c
{\cal L}^{res}_{tH}(2)$ has the form [11, 12]
%8
$${\cal L}^{res}_{tH}(2) = G^{tH}\{~[(\bar{q}_iq_i)(\bar{q}_fq_f) -
(\bar{q}_i\tau q_i)(\bar{q}_f\tau q_f)~] -$$
\be
-[~(\bar{q}_i i\gamma_5q_i)(\bar{q}_f i\gamma_5 q_f) - (\bar{q}_i
i\gamma_5\tau q_i)(\bar{q}_f i\gamma_5\tau q_f)~]~\}
\ee
where  $\tau$ are Pauli matrices.

The transition $\bar{u} u \to \bar{d} d$ corresponds to $180^o$ rotation of
$\bar{u}$ and $u$ around the second axis in the isotopical spin space
($u \to -d,~ \bar{u} \to -\bar{d}$) [17]. Since at such rotation bilinears
of scalars and pseudoscalars go over into each another we restrict ourselves
when analysing reactions 1 and 2, to the term in (8) averaged over isospin
%9
\be
{\cal L}^{res}_{tH}(\bar{u}u \to \bar{d}d) = G [~(\bar{u}u)(\bar{d}d) -
(\bar{u}i\gamma_5u)(\bar{d}i\gamma_5 d)~].
\ee
$\bar{u} u (\bar{d} d)$ are superpositions of states with $I = 0,1$ that
is why in intermediate two-quark states we fix the third component of
isospin $I_3 = 0$, but not $I$.

The extended NJL model does not consider the problem on conservation of
$l_z, S_z$ in $(\bar{u}u)(\bar{d}d)$, $(\bar{u}u)(s\bar{s})$.

In order that to turn from contact amplitudes to amplitudes (4) one should
take in to account propagators of scalar and pseudoscalar states $D_{S,
PS}$. To this end the authors of [11,12] used random phase approximation
[18]
%10
\be
D_{S,PS} =
\frac{2iG}{1 -2 G\Pi_{S,PS}}
\ee
where $\Pi _{S, PS}$ are loop integrals
depending on the total energy of charge-exchanged quarks $k^2 = 4 m^2_q + 4
\vec{k}^2$, on the intermediate state mass $m^*_{S, PS}$ and $\Lambda$.

Note three, important for our following analysis, consequences of the NJL
model:

1. It is shown in [11] (eqs.4.14 - 4.32) and furthermore 6.49) that
%11
\be
1 - 2G\Pi_{S,PS} = \frac{m}{m_q} + i2G(k^2 - m^{*2}_{S,PS})I^{\prime}(k^2,
m_q, \Lambda)
\ee
where $I^{\prime}$ -- a is fluent, weakly depending on $k^2$ function.
Substituting (11) into (10) we see that after spontaneous violation of
chiral symmetry (i.e. at $G \geq G_c,~` \frac{m}{m_q} = \frac{m}{M_c} \simeq
0, ~~ D_{S,PS}$ stops depending on $G$ provided that $k^2 \not= m^{* 2}
_{S,PS}$.

G-independence of $D_{S,PS}$ at $G \geq G_c$  suggests that as
constituent quarks at the edges of two f.t. appear to be in $^3P_0$ or
$^1S_0$  states they annihilate with a unity probability (two f.t. fuse
into one with a unity probability):
%12
\be
\gamma^2~~ \footnote{The value $\gamma^2$ was formulated by the author
  together with L.B.Okun in [6].} =  \frac{\mbox{number of annihilations
with production of general colour f.t.} c^{-1}} {\mbox{number of
colour}~^3P_0(~^1S_0)\mbox{pairs} \cdot c^{-1}} = 1 \ee We call the equality
$\gamma^2 = 1$ as "blackness" condition.  The "blackness" condition (12)
will be used to estimate the absolute value of $M_1$ in the 4-th section.
The "blackness" effect is a consequence of spontaneous violation of chiral
symmetry. Note that G-independence of amplitudes at $G \geq G_c$ by no means
implies that in $T(S)$ and $T(PS)$ (4) the values $G_S$ and $G_{PS}$ are
arbitrary. In fact, $G_S = G_{PS} = G$ since (11) follows from (5).

2. In the NJL model $m^*_S \equiv m_{\sigma} =
2M_c$ (see experimental confirmation in \cite{25} and references therein. 
At $G \geq G_c $
%13
\be
k^2 = 4M^2_c + 4\vec{k}^2 .
\ee

Substituting $ m^2_{\sigma}$ into (11) we get (accounting for (13)
and (10)) ${D_S = (\vec{k}^2 \cdot I^{\prime})^{-1}}$. Thus, $D_S$ in (4)
is determined by the 3-moment  of charge-exchanged quarks , i.e. by the
dimension of the region where charge exchange occurs. In the 4-th section we
shall see that the transversal part of this region ${R^2 = \langle
k^2_{\bot} \rangle^{-1}}$ enters the expression for the amplitude $M$ as
the radius of the residue of the Regge pole M which dominates in the
reaction.

3. Now we can write in (4):
%14
$$T(S) - T(PS) \approx \frac{(\bar{u}u)(\bar{d}d)}{k^2 - m^2_{\sigma}} -
\frac{(\bar{u}i\gamma_5 u)(\bar{d}i\gamma_5 d)}{k^2 - m^{*2}_{PS}} =$$
\be
\frac{(\bar{u}u)(\bar{d}d)}{4\vec{k}^2} - \frac{
(\bar{u}i\gamma_5 u)(\bar{d}i\gamma_5 d)}
{4M^2_c + 4\vec{k}^2 - m^{*2}_{PS}}
\ee
Since $(\bar{u}u)(\bar{d}d)=4\vec{k}^2$;
$(\bar{u}i\gamma_5 u)(\bar{d}i\gamma_5 d) = 4M^2_c + 4\vec{k}^2$
%15
\be
T(S) - T(PS) \approx 1 - \frac{1}{1 - \frac{m^{*2}_{PS}}{4M^2_c +
4\vec{k}^2}}
\ee

In the minimal NJL model $m^*_{PS} = m_{\pi}$. The amplitude $M_1$ contains
only $M_1(^3P_0)$, while $M_0$ -- the combination of $M_0(^3P_0)$ and $M_0(^1S_0)$
(see the next section). Because of (15) the NJL model predicts a
strong destructive interference of $M_0(^3P_0)$ and $M_0(^1S_0)$ in $M_0$.

\section{Analysis of experimental data of reactions $\pi^-p \to
\pi^0n; ~~ \pi^-p \to \eta n; ~~ \pi^- p \to K^0 \Lambda(\Sigma)$.}
I. $\pi^-p \to \pi^0 n, ~~ \eta n$.

In papers of the High Energy Physics Institute -CERN [1,2] the world data on
differential cross sections of reactions are well parametrized in the
region $s = 8-400 GeV^2$ and $q^2_{\bot} = 0.004-0.3(GeV/c)^2$ as
%16
\be
\frac{d\sigma}{dq^2_{\perp}} =
\frac{d\sigma}{dq^2_{\perp}}\vert_{q_{\perp}=0} (1 + aq^2_{\perp})exp
[-2q^2_{\perp}(R^2 + \xi \alpha^{\prime})]
\ee
$$(d\sigma/dq^2_{\perp})_{q_{\perp}=0}=A~exp(-2\beta\xi);~~\xi =
ln\frac{s}{s_0};~~ ~s_0 = 1 GeV^2 \simeq (\alpha^{\prime})^{-1}$$

The parameters of (16) are listed in Table 1.

\begin{table}
%Table 1
\addtocounter{footnote}{1}
\caption{$^\thefootnote$
The list of parameters (16), describing differential cross sections of
reactions 1 and  2 within the interval 
$s = 8 - 400 GeV^2, q^2_{\perp} = 0,004 - 0,3 (GeV/c)^2$ 
[1,2 Apel V.D.],
$r^* = [1 + \rho^2(0)]^\frac{1}{2}$}

\begin{center}
\begin{tabular}{|l|l|l|l|l|l|l|} \hline
& ~~ $R^2$ & $a$  & $A$ & ~~~$\alpha^{\prime}$ &
$\beta=1 - \alpha(0)$ &~ $r^*$ \\
& (GeV/c)$^{-2}$
& (GeV/c)$^{-2}$ & (GeV)$^{-4}$ & (GeV/c)$^{-2}$  && \\ \hline
&&&&&& \\
$ \pi^0 n$ & $4,5(1 \pm
0,03)$ & $33(1 \pm 0,05)$ & $12(1\pm 0,05)$ & $0,8(1 \pm 0,06)$& $0,52(1 \pm
0,02)$ & 1,37 \\
&&&&&& \\
$\eta n$ & $1,16(1 \pm 0,09)$ & $31(1
\pm 0,2)$ & $1,49(1 \pm 0,1)$ & $0,8(1 \pm 0,06)$ & $0,603(1 \pm 0,03)$ &
1,70 \\
&&&&&&\\ 
\hline
\end{tabular}
\end{center}

{$^\thefootnote$ \footnotesize Parameters are taken from [1] Apel V.D.
et al. Yad.Fiz. V.30, p.373 and [2] Apel V.D. et al. Yad.Fiz. 1979, V.29,
p.1519. The parameters $R^2$ and $A$ depend on the value of dimensional
factor $s_0$  in $\xi$. In the papers by Apel et al. it was adopted $s_0 =
10 GeV^2$. In this paper $s_0 = 1 GeV^2 \simeq (\alpha^{\prime})^{-1}$.
Parameters $R$ (Apel) and $A$ (Apel) were recalculated taking into
account $s_0 = 1 GeV$. When recalculating $R^2$, the absolute values of
errors were conserved in $R^2$ (Apel) and in $\alpha^{\prime}$. When
recalculating $A$ relative errors $A$ (Apel) were conserved.
In papers by V.D.Apel et al. $(a)$
were parametrized by the functions $a_1 = [12,7(1 \pm 0,024) + 1,57(1\pm
0,076) \ln \frac{s}{10}] \cdot [2,55 (1 \pm 0,035) - 0,23(1 \pm 0,26)ln 
\frac{s}{10}]$; $a_2 = [6,0(1 \pm 0,03) + 1,6(1\pm 0,06) \ln
\frac{s}{10}] \times [4,6(1\pm 0,07) - 0,5(1\pm 0,4) \ln\frac{s}{10}]$.
$a_{1,2}$  weakly depend on $s$ in the interval $s = 8 - 400 GeV^2$.
Table 1 shows means arithmetics $a_{1,2}$ taken at
$s=10, 100, 400 GeV^2$.}
\end{table}

\vspace{1mm}
We assume that
%17
\be
a = \frac{\mid M_1 \mid^2}{\mid M_0 \mid^2}~\frac{R^2}{4}
\ee
Whence
%18
\be
\mid M_1 \mid /\mid M_0 \mid \equiv \nu = 5.4; 10.3
\ee

For reactions 1 and 2, correspondingly. The assumption (17) is confirmed by
the following data:

1. A direct amplitude analysis of $\pi N \to \pi N$ including data on spin
flip at $s \cong 12 GeV^2$ in the region $q^2_{\bot}\leq 0.5
(\frac{GeV}{c})^2$ gives for the charge exchange amplitude $M_1/M_0 = -8$ in
accordance with the $M_{0,1}$ calculations on the basis of dispersion sum
rules at finite energy [19,20].

2. The phase analysis $\pi N \to \pi N$ at $P_{lab} = 2 GeV/c$ gives
for the charge exchange amplitude $M_1/M_0 = -4 \div 5$ [21]. $R^2$ at the
points 1 and 2 is taken from the data on $\pi^-p \to \pi^0 n$ at $s = 8 -400
GeV^2$.

3.Polarization in the region $q^2_{\bot} \leq 0.3(GeV/c)^2$ and $s = 80
GeV^2$ is small [22], so that the relative phase of $M_0$ and $M_1$ is
$\vert \varphi_0 - \varphi_1 \vert \leq 4.10^{-2}$ for both reactions. In
what follows we will neglect  $\vert \varphi_0 - \varphi_1 \vert$ and will
consider $M_0/M_1$ to a real quantity.

4. (16) coincides with the Regge expression for reactions $0^- \frac{1}{2}
\to 0^- \frac{1}{2}$ with one pole in $t$ channel and equal residue radius
$R$ in $M_0$ and $M_1$.

5. In the next section eq.(17) will be quantitatively confirmed.

If we restrict ourselves in $s$ channel  of the dual diagram of Fig.3b to
$S$ and $P$ combinations of charged exchanged quarks (minimum  sufficient
for self-consistent description of reactions 1 and 2), then the subprocess
$\bar{u} u \to \bar{d} d$ may involve the following states:

\vspace{3mm}
\begin{center}
\begin{tabular}{lllll}
 $S$(scalar) &
~~~~~$V$ & ~~~$~T$ & ~~~$~A$ & ~~~$~P$ \\
&&&& \\
$^3P_0$  & $^3S_1, ~^1P_1$
& ~~~$^3P_2$ & ~~~$^3P_1$ & ~~~$^1S_0$ \\
\end{tabular}
\end{center}

\vspace{3mm}
Only $^3P_0$ and $^3P_2$ may contribute to $M_1$ since spin-orbital amplitude
in the subprocess appears if there is a difference in cross sections in pair
annihilation with $l_z = \mp 1$ at $S_z = +1$. Sign inversion in $l_z$
transfers $^3P_0 (^3P_2)$ into another state -- $^3P_2(^3P_0)$. In the rest
cases there are no changes, as either $l_z = 0$ or $S_z = 0$.

$M_0$ can be , in principle, contributed by all 6 states.

In the case of the $^3P_0$ or $^3P_2$ dominance and at $q^2_{\bot} R^2 \ll
1$ $\vert M_2 (^3P_{0,2}) / \vert M_0 (^3P_{0,2}) \vert = 2$ [6], see also
(38). Hence it follows that the experimental values $\nu = 5; 10$ (18)
result from destructive interference of $M_0(^3P_{0,2})$ with other states
[6]:
%19
\be
\nu = \frac{\mid M_1 \mid}{\mid M_0 \mid} =
\frac{M_1(^3P_{02})\mid}{\mid M_0(^3P_{02}) - M_0(x)\mid } = \frac{2\mid
M_0(^3P_{02})\mid}{\mid M_0(^3P_{02}) - M_0(x)\mid},
\ee
where $x=^3S_1$, $^1P_1$, $^3P_1$ and $^1S_0$.

In Table 2 we present the Fiertz expansion of the soft "t"-channel vector
($\rho$-exchange in reaction 1) and tensor ($A_2$ exchange in
reaction 2)\\ 4-quark amplitude $\bar{u} d \to \bar{u} d ~ (\bar{q}_1 q_3 \to
\bar{q}_2 q_4$ in Fig.3b) over $s$-channel amplitudes $\bar{u} u \to
\bar{d}d$ of interest.

\begin{table}
%Table 2.
\caption{Fiertz expansion of the soft $t$-channel amplitude 
$\bar{u}d \to \bar{u}d$
over $s$-channel amplitudes $\bar{u}u \to \bar{d}d$. Only bilinear notations
are presented.}

\begin{center}
\begin{tabular}{|l|ccccc|} \hline
\multicolumn{1}{|l|}{t \mbox{channel}} & \multicolumn{5}{|c|}{$s$~~
\mbox{channel}~~ $\bar{u}u \to \bar{d}d$}\\
$\bar{u}d\to \bar{u}d$ & $S$ &
$V$ & $T$ & $A$ & $PS$ \\  \hline
&&&&& \\
$\rho$ \mbox{exchange}~$(^3S_1)^2$ & ~~~$(^3P_0)^2$ &
~~~$(^3S_1)^2$ & ~~--  & ~~~$(^3P_1)^2$ & ~~~$(^1S_0)^2$ \\
&&&&& \\
$A_2$
\mbox{exchange}~$(^3P_2)^2$ &  ~~~$(^3P_0)^2$ & ~~~-- & $(^3P_2)^2$ &
~~~-- & ~~~$(^1S_0)^2$  \\ \hline
\end{tabular}
\end{center}
\end{table}

"s"-channel spin structures of the amplitudes $\bar{u} u \to \bar{d} d$ with
the opposite $"t"$ channel signature ($\rho$ and $A_2$) differ by their
$"u"$ channel singularities. But in $"u"$ channel ($\bar{u} \bar{d} \to
\bar{u} \bar{d})$ repulsion forces predominate and there are no physical
 poles; for this reason $"u"$ channel gives no large contribution
  into $"s"$ channel that is confirmed by exchange degeneration of $\rho$ and
$A_2$ trajectories. This argumentation allows one to narrow the set of $"s"$
  channel bilinears of Table 2 in reactions 1 and 2 up to two general
$(^3P_0)^2 = (\bar{q} q)^2$ and $(^1S_0)^2 = (\bar{q} i \gamma_5 q)^2$.
  Thus, reactions 1 and 2 involve the same bilinear forms as in the
minimal NJL model. Now eq.(19) is open:
%20
\be
\nu = \frac{2 \vert M_0 (^3P_0) \vert}{\vert M_0(^3P_0) - M_0(^1S_0) \vert}
\ee
From (20) we determine the relative contributions

$M_0(^3P_0)$  and $M_0(^1S_0)$:  $\chi(^3P_0) = \mid M_0(^3P_0)\mid$ $/
\mid M_0(^3P_0)\mid + \mid M_0(^1S_0)\mid$; $\chi(^1S_0) = 1 - \chi(^3P_0)$
%21
\be
\chi(^3P_0) = \frac{\nu}{2(\nu -1)}
\ee
see Table 3.

\begin{table}
%Table 3.
\caption{Relative contributions $\mid M_0 (^3P_0)\mid$  and $\mid
M_0(^3S_1)\mid$  in reactions 1 and 2.}

\begin{center}
\begin{tabular}{|l|l|l|l|l|}  \hline
& $\nu = \mid M_1 \mid / \mid M_0 \mid$ & $\chi(^3P_0)$ & $\chi(^1S_0)$&
\mbox{interference}~~$M_0(^3P_0), M_0(^1S_0)$\\
& &&& \\ \hline
$\pi^- p \to \pi^0 n$ & ~~~~~~~~~5.4 & ~~0,6 & ~~0,4 &
~~~~~~~~~\mbox{destructive} \\
&&&& \\
$\pi^- p \to \eta n$ &  ~~~~~~~~~10.3 &
~~0,55 & ~~0,45 & ~~~~~~~~~\mbox{destructive} \\  \hline
\end{tabular}
\end{center}
\end{table}

The data of Table 3 agree with predictions of the extended NJL model --
destructive interference of $S$ and $PS$ bilinears in $M_0$ and their close
absolute values (relic of equality $G_S = G_{PS} = G$ in the bare Lagrangian
of NJL (5).

\vspace{3mm}
II. Reactions with strangeness production of $\pi^-p \to K^o
\Lambda(\Sigma)$.\\
Besides the differential cross sections data of Fig.3,
there are the data [3] (B.Foley et al.) in the region $s = 16 - 30 GeV^2$
and $q^2 _{\bot} = 0 \div 1 (GeV/c)^2$ where spin-orbital dip is also
invisible, but the errors in $(d \sigma/ dq^2_{\bot})_{q_{\bot} = 0}$ are
presented and $\alpha^{\prime}_S$ is estimated. From these data one can
estimate the upper limit for the value $\Biggl ( \frac{\vert M_1 \vert}{\vert
M_0 \vert} \Biggr )_S$ according to formula
%22
$$
\frac{d\sigma}{d q^2_{\perp}} = \Biggl (\frac{d\sigma}{d q^2_{\perp}}
\Biggr )_{q_{\perp}=0} ( 1  + \delta) exp [-2q^2_{\perp}(R^2 +
\alpha^{\prime}_s\xi) =$$

\be
= \Biggl ( \frac{d\sigma}{d q^2_{\perp}} \Biggr )_{q_{\perp}=0}
(1 + \Biggl ( \frac{\mid M_1 \mid^2}{\mid M_0 \mid^2}\Biggr )_s
\frac{R^2}{4}\cdot \Delta q^2_{\perp}\Biggr ) exp [-2q^2_{\bot}(R^2 +
\alpha^{\prime}_s\xi])
\ee
where $\delta = 0.022$ is an error in $d \sigma/d q^2_{\bot}$ at the minimal
$q^2_{\bot}$ within the minimal interval $\Delta q^2 = 0.- 0.05
(GeV/c)^2,~~ R^2 = 2.8 (GeV/c)^2$ and $\alpha^{\prime}_s = 0.43 \pm 0.36
(\frac{geV}{c})^{-2}$. As a result:

%23
\be
\Biggl ( \frac{\mid M_1 \mid}{\mid M_0 \mid}\Biggr )_s  \leq 1.
\ee

${\cal L}^{res}_{tH}(4)$ (7) is formally $SU(3)$ symmetrical. What is the
reason for strong $SU(3)$ violation in $M_1$ at production of strangeness?
A possible answer is that in the instanton transition $\bar{u}u \to s
\bar{s}$ [23], the NJL model does not take into
account the $l_z$ non conservation in $(\bar{u}u)(s \bar{s})$. In this
case $(\bar{u}u)(s \bar{s}) = (^3P_0)^2$ does not contribute to the amplitude
$M$ in "q" representation (see the next section).

\section{Calculation of the absolute value of $M_1$ in reactions 1,2.
Comparison with experimental data.}
Thus, $M_0 = M_0(^3P_0) - M_0(^1S_0)$ is a small difference of large
quantities and, consequently, is theoretically unstable. The instability is,
due, in particular, to a different contribution of Regge cuts $R \otimes P$
in $M_0(^3P_0)$ and $M_0(^1S_0)$ [14]. While $M_0(^3P_0)$ does not
practically contain  $R \otimes P$ [24], the contribution from $R \otimes P$
into $M_0(^1S_0)$ diminishes it by $\simeq 20\%$ [14]. That is why we
restrict ourselves to calculation of the dominant $M_1$, \footnote{ The part
of $M_1$ in the total cross sections of reactions 1 and 2 comprises $\simeq
90\%$ at $P_{lab} \simeq 10 GeV/c$} which within the model equals to
$M_1(^3P_0)$.

We compare $\vert M_1(^3P_0) \vert$ with $\vert M_1(\mbox{exper.}) \vert$.
From (1), (16), (17)

%24
\be
\mid M_1(\mbox{exper.}) \mid =\mid M_1(\mbox{exper.})\mid
_{q_{\bot}=0} \cdot exp[-q^2_{\perp}(R^2 +
\alpha^{\prime}\xi)]
\ee
%25
\be
\vert M_1 (\mbox{exper} \vert_{q_{\bot}=0} = 8 \sqrt {\pi} p_c \sqrt {s}
\frac{2 \sqrt {a}}{R}~A^{1/2} exp (-\beta \xi)
\ee
(the parameters (24) and (25) in Table 1).

To calculate $\vert M_1(^3P_0) \vert$ we need the unitarity conditions of
f.t.
model of binary processes. Multiple processes do not enter the unitarity
conditions of f.t. model since they occur after the first disruption of f.t.
and causally is almost independent of the binary process. The intermediate
stage contain only  single f.t. which may differ only by transformation
properties of the first disruption mode. The unitarity conditions in $s$
channel in $"b"$ representation have the form [6]
%26
\be
Im~ M_{\bf b} (ab \to cd) = 2 p_c  \sqrt{s} \sum \limits ^{5}_{i = 1}~
M_{\bf b} (ab \to f.t._i) M^*_{\bf b} (f.t._i \to cd)
\ee
where
%27
\be
\vert M_{\bf b}(ab \to f.t.) \vert^2 = W_{\bf b}(ab \to f.t.) = N_c W_{\bf
b} \Biggl [ (ab)_c \to f.t._c \Biggr ]
\ee

$W_{\bf b} \Biggl [(ab)_c \to f.t._c \Biggr ]$ is the probability of
production of a coloured f.t. at annihilation of a coloured $\bar{q} q$
pair. Since we estimate only $^3P_0$ mode of disruption, index $i$ in (27)
and hereafter is omitted.

Up to exchange degeneration of reactions $\pi^- p \to f.t. \to \pi^- p$ and
$\pi^0 n \to f.t. \to \pi^0 n$ where  only contributions from the $\rho$ and
$f$ pole, respectively,  are taken into account

%28
$$
(2 p_c \sqrt{s})^{-1} \vert Im M_{\bf b} (\pi^- p \to \pi^0 n) \vert = \Biggl
[\vert M_{\rho} (\pi^- p \to f.t.) \vert^2 \vert M_f (\pi^0 n \to f.t.) \vert
^2 \Biggr ]^{1/2} =
$$
\be
= \vert M_{\rho} (\pi^- p \to f.t.) \vert ^2
\ee
From (28) and (27)

$$
(2 p_c \sqrt{s})^{-1} \vert Im M_b(\pi^- p \to \pi^0 n) \vert = N_c W ({\bf 
b}, y_d - y_{ud}) = $$ 
%29 
$$ N_c \int~ d^2 ({\bf b}_u - {\bf b_{\bar{u}}}) 
d^2 ({\bf b}_d - {\bf b}_{\bar{u}})\omega( -y_d, {\bf b}_d - {\bf b}_{\bar{u}}) 
W ({\bf b}_u - {\bf b}_{\bar{u}}) \omega (y_{ud}, {\bf b}_{ud} - 
{\bf b}_u) \equiv $$ 
\be 
\equiv < f.t. \vert T (^3P_0) \vert f.t. > 
\ee 
where $W ({\bf b}, y_d- y_{ud})$ is a dimensionless probability of production
of a coloured f.t. with coloured quark and diquark at the edges with the
rapidity difference $y_d - y_{ud} = \xi$ and with projection ${\bf b} =
{\bf b}_d - {\bf b}_{ud}$ onto the plane of the impact parameters $y, z$
(Fig.3d);
%30
\be
\omega(y_i, {\bf b}_i - {\bf b}_k) = \beta(4 \pi \alpha^{\prime} y_i)^{-1}
exp \Biggl [-\beta y_i - ({\bf b}_i - {\bf b}_k)^2 (4 \alpha^{\prime}
y_i)^{-1} \Biggr ]
\ee
$$
\beta = 1 - \alpha_{\rho}(0)
$$

is a normalized [6] density of a probability to find in the impact parameter
plane a parton at the point ${\bf b}_k$ with rapidity $\sim 0$ if it comes
out from  the point ${\bf b}_i$ with rapidity $y_i$ [9].

The dimensionless probability of production of a unite coloured f.t. at
annihilation of a coloured pair from $^3P_0$ state:
%31
\be
W({\bf b}_u - {\bf b}_{\bar{u}}) \approx \vert \int~ d^2 k_{\bot}
dk_{\parallel} exp [-2R^2(k^2_{\parallel} + k^2_{\bot})]
j^c_S ({\bf k}) exp \Biggl [-i {\bf k} ({\bf b}_u - {\bf b}_{\bar{u}})
\Biggr ] \vert^2
\ee
is the square of the Fourier-image of scalar current with transformation
properties $^3P_0 = J^P = O^+$. In the given model the current $J^c_S$
corresponds to annihilation (production) of a constituent pair
of converging (diverging) waves with moments ${\bf k}$ and $-{\bf k}$ and
with unit summary spin and orbital moment with conserved $S_Z = +1$ and $l_Z
= -1$, correspondingly:
$$
j^c_S = (8 \pi)^{1/2} ~\sum_{m = 0,1}~ C(110; m, -m, 0) \times
C(\frac{1}{2}, \frac{1}{2}, 1; m_1, m_2, m) Y_{1, -m} ({\bf k}) \chi_{m_1}
\chi_{m_2} =
$$
%32
\be
2 [ i(k_x - ik_y) \chi_{1+1} - ik_z \chi_{10} ]
\ee
where $\chi_{1,m}$ are spin functions of a pair. The parameter $R^2$
determines the scale of annihilating quark momenta. It enters the final
expression of the amplitude as the radius of the Regge pole residue.
Integration over $k_{\parallel} = k_x$ is performed within $- \infty <
k_{\parallel} < 0$ (converging wave) Integration over $d k_{\parallel}$ and
$d^2 k_{\bot}$ and squaring results in:

$$
W({\bf b}_u - {\bf b}_{\bar{u}}) = W ({\bf b}^{\prime}) =
$$
%33
\be
= \gamma^2 \cdot 0.464 (1 + 1.25 \vert {\bf b}^{\prime} \vert R^{-1}~ Cos~
\varphi_{b^{\prime}} + 0.39 b^{\prime 2} R^{-2} )exp \Biggl
(-\frac{b^{\prime 2}}{4 R^2} \Biggr) = \gamma^2 \cdot F({\bf b}^{\prime})
\ee
where $\varphi_{b^{\prime}}$ is counted from the axis $Y$. $F({\bf
b}^{\prime})$ -- the ${\bf b}^{\prime}$-dependence of  the pair production
probability in $^3P_0$ state is determined  such that $0 \leq F \leq 1$.
$\gamma^2 = 1$ in
accordance with the blackness condition (12).\footnote{In [6] (expression 20) the author had obtained the correct value
$\gamma^2_1 \simeq 1 ( \gamma^2_1 = \gamma^2$ of this paper).
Unfortunately, $\gamma^2_1 \simeq 1$ was obtained as a result of a random
compensation of two errors: the numerator  in (20) [6] did not take into
account the factor $N_c = 3$ in front of $\gamma^2$. The denominator in (20)
[6] used $\sqrt{A(Apel)}$ instead of $\sqrt{A}$ from Table 1 of the given
work (see footnote 2) and, consequently, factor 3.3 and 4 for reactions 1
and 2, respectively, was not taken into account}.

The scalar current $j^{nc}_S$ with transformation properties $J^P =
^3P_0 = 0^+$ but not conserving $l_z, S_Z$ gives no contribution into
amplitude $M(q^2_{\bot})$ (1) since in
%34
$$
[(q_i q_i) (q_f q_f)]^{nc} \approx
$$
\be
\approx (Y_{1-1} \chi_{1+1} + Y_{10} \chi_{10}) (Y^*_{1+1} \chi_{1-1} + 
Y_{10} \chi_{10}) 
\ee 
$Y_{1-1} Y^*_{1+1} \chi_{1+1} \chi_{1-1} = 0$ by 
virtue of orthogonality, while $(\chi_{10} Y_{10})^2$ is out when going over 
into $"q_{\bot}"$ representation [14].

After substituting (30) and (33) the
convolution (29) is integrated. As a result we obtain the one-pole
charge-exchange amplitude where only $T(S)$ in (4) is taken into account:

%35
$$
(2p_c) \sqrt{s})^{-1}~ Im M_b (^3P_0) =
$$
\be
\frac{G^{\prime}}{C^{\prime}} \Biggl
(1 + 1.56 \frac{C^{\prime} - R^2}{C^{\prime}} + 0.39 \frac{R^2
b^2}{C^{\prime 2}} + 1.25 \frac{\vert{\bf b} \vert R}{C^{\prime}} \cdot Cos
\varphi_b \Biggr ) exp \Biggl (-\frac{b^2}{4 C^{\prime}} \Biggr )
\ee
where $P_c \vert {\bf b} \vert Cos \varphi_b = -{\bf l} \vec{\sigma},
C^{\prime} = R^2 + \alpha^{\prime} \xi;~~~ G^{\prime} = 0.464 N_c \beta^2
R^2 exp(-\beta \xi)$; at $\xi = 0$ (35) goes over into (33) up to a
coefficient $N_c \beta^2 R^2$.

In $"q"$ representation $(Im M(q_{\bot}) = \int~ Im M_b~ exp(-i q b) d^2 b)$
%36
\be
Im~ M(^3P_0) = Im~ M_0 + i \vec{\sigma} {\bf n} ~ Im M_1 \frac{q_{\bot}
R}{2}
\ee

%37
\be
Im~ M_0(^3P_0) = + G \cdot exp(-C)
\ee

%38
\be
Im~ M_1(^3P_0) = -1.95~ G \cdot exp(-C)
\ee
where $G = -2 p_c \sqrt{s} \cdot 2 \pi R^2 N_c \beta^2 \cdot 0.464 \cdot
5.12$ at $q^2_{\bot} R^2 \ll 1; \\ C = q^2_{\bot} (R^2 + \alpha^{\prime}
\xi) + \beta \xi$.

For the choice of signs of $Im~ M_{01}$ and $G$ see [6].

Thus:

$$
\vert M_1 (^3P_0) \vert = [1 + \rho^2(q_{\bot})]^{1/2} \vert Im~M_1(^3P_0)
\vert = [1 + \rho^2(q_{\bot}) ]^{1/2} \cdot 2 \pi N_c R^2 \beta^2 \times $$
%39
\be
\times 1.95 \cdot 0.464 \cdot 5.12 \cdot 2 p_c \sqrt{s}~ exp (-C)
\ee
where $\rho(q_{\bot}) = tg \frac{\pi}{2} [\alpha_{\rho}(0) +
\alpha^{\prime} q^2_{\bot} ];~~ ctg \frac{\pi}{2} [\alpha_{A_2}(0) +
\alpha^{\prime} q^2_{\bot} ]$ for reactions 1 and 2, respectively.

We get from (39) and (24):
%40
\be
\frac{\vert M_1(^3P_0) \vert}{\vert M_1(exper.) \vert} =
\frac{[1 + \rho^2 (q_{\bot}) ]^{1/2} \cdot 2\pi N_c R^2 \beta^2 \cdot 4.63
\cdot 2p_c \sqrt{s} \cdot exp(-C)}{8 R^{-1} \sqrt{\pi a A}
\cdot 2p_c \sqrt{s}~ exp(-C)}
\ee
Substituting into eq.(40) the parameters $R, a, A, \beta, [1 +
\rho^2(0)]^{1/2}$ from Table 1
we have
%41
\be
\frac{\vert M_1(^3P_0) \vert}{\vert M_1(exper.) \vert} = 1.09(1 \pm 0.1);~~~
0.7(1 \pm 0.2)
\ee
for reactions $\pi^-p \to \pi^0 n$ and $\pi^-p \to \eta n$, correspondingly.
From (41) it follows, that (39) describes, within the errors of existing
experiments, $\vert M_1(exper) \vert$ in reactions 1  and 2 in the interval,
$s = 8 - 400 GeV^2$ and $0.004 \leq q^2_{\bot} \leq 0.1 (\frac{GeV}{c})^2$.

\section{Conclusion}
Charge-exchange reactions $\pi^-p \to \pi^0 n,~ \eta n; ~ \pi^- p \to K^0
\Lambda(\Sigma)$ occur through two phases.

1. Diffusion of constituent quarks of each interacting hadrons in the space
of rapidities and impact parameters involving flux-tube production, one edge
of this f.t. having fast spectator and another -- slow reagent. Confinement
forces ($\alpha^{\prime}$ in the diffusion equation (31)) are explicitly
taken into account in this phase); it determines the Regge power
decrease of the amplitude with energy increase.

2. Charge-exchange of slow  reagents: $\bar{u} u \to \bar{d} d, ~~ \bar{u}
u \to s \bar{s}$. This phase determines the spin structure and the absolute
value of the Regge pole residue.

The spin-orbital amplitude $M_1$
, caused by
spin-orbital interaction (SOI) in the process $\bar{u} u \to \bar{d} d$,
dominates in reactions $\pi^- p \to \pi^0 n, ~ \eta n$ (more than $\%90$ of
cross section  of the reaction). SOI stems from quark charge-exchange in
the state $J^{\pi} = 0^+ =^3P_0$ provided $l_z$ conservation in this
process.  Consequently, the process $\bar{u} u \to \bar{d} d$ is planary
(Fig.4a).  This fact is an argument in the favour of that to consider the
simplest planary dual diagram of Fig.3b as an experimentally substantiated
model of reactions.

The NJL model gives certain predictions on relations between different
4-quark invariants in $\bar{u} u \to \bar{d} d$ and, consequently, in
charge-exchange physical amplitudes. In the minimal NJL model only bilinears
of scalars $(^3P_0)^2$ and of pseudoscalars $(^1S_0)^2$ are effective. They
enter the physical non-flip amplitude $M_0$ with the opposite signs and with
approximately equal weights (strong destructive interference). The weights
are independent of the coupling constant of NJL Lagrangian (the
"blackness" condition (12)) and are determined by the dimensions of the
region of charge exchange $\bar{u} u \to \bar{d} d$. The transversal radius
of this region $R^2$ enters the finite expression of the physical amplitude
as a radius of the Regge pole residue and determines its absolute value.
These predictions are direct consequences of dynamical violation of the
chiral invariance and of non conservation of the $SU(3)$ singlet current
, underlying the extended NJL model.

Luckily, the most completely experimentally studied reactions $\pi^- p \to
\pi^0 n (\eta n)$ contain the same bilinears $(^3P_0)^2$ and $(^1S_0)^2$ as
the minimal NJL model. This conclusion resulted from comparison of the
Fiertz expansion of the t-channel soft 4-fermion amplitude in reactions
$\pi^- p \to \pi^0 n (\eta n)$ with the account of their exchange
degeneration.  The analysis of experimental data had confirmed the
consequences of the NJL model - the strong destructive interference $^3P_0$
and $^1S_0$ in $M_0$. We have refrained from estimating the absolute value
of $M_0$ in view of its theoretical instability (a small difference of large
values).

The absolute value of $M_1~ (^3P_0)$ contain; $R^2$, the colour number $N_c
= 3$ and $\beta^2 = (1 - \alpha_R(0)^2$ and, consequently, there are
no free parameters. The comparison with the experimental $M_1(exper.)$
yields:

$$
\vert M_1(^3P_0) \vert / \vert M_1(exper.) \vert = 1.09(1 \pm0.1);~~ 0.7(1
\pm 0.2)
$$
for reactions $\pi^- p \to \pi^0 n (\eta n)$, correspondingly, within the
interval $s = 8 - 400 GeV^2$ and $0.004 \lesssim q^2_{\bot} \lesssim
0.1(\frac{GeV}{c})^2$.

In reactions $\pi^-p \to K^0 \Lambda (\Sigma)$ SOI are at least by the order
of magnitude smaller than in $\pi^- p \to \pi^0 n (\eta n)$. The transition
$\bar{u} u \to s \bar{s}$ is usually considered as an instanton one. On the
other hand, in the $SU(3)$ symmetrical model NJL, the process $\bar{u} u \to
s \bar{s}$ also contains the scalar bilinear. The absence of SOI in $\pi^- p
\to K^0 \Lambda(\Sigma)$ may be, perhaps explained as follows: $l_z$ is not
conserved in the instanton transition $\bar{u} u \to s \bar{s}$ in $^3P_0$
state. In this case $(\bar{u} u)(\bar{s} s)$ gives no contribution into
physical amplitude M (see the 4-th section). Thus, a strong $SU(3)$
violation appears in the formal $SU(3)$ symmetrical NJL model.

A question arises why in the scalar channel of the soft instanton
transition $\bar{u} u \to \bar{d} d ~~ l_z$ is conserved, while
in $\bar{u} u \to s \bar{s}$ it is not?

Reactions $\pi^- p_{\uparrow} \to K \Lambda_{\nearrow}(\Sigma_{\nearrow})$
are very interesting objects for studying transitions $\bar{u} u \to s
\bar{s}$ since the non conserving parity decays $\Lambda_{\nearrow}
(\Sigma_{\nearrow}) \to \pi^- p$ provide an experimental possibility for the
total and exact amplitude analysis of these reactions in the whole region of
$q_{\bot}$.

In this work we based on: 1. The parton concept of binary processes. 2.
 Non conservation of the $SU(3)$ singlet current (gluon anomaly consequence).
3.The "blackness" condition which follows from dynamical violation of chiral
symmetry. In view of a general character of these elements, the above
mentioned give, as it seems to us, grounds to believe soft binary processes
as a good laboratory of binary quark reactions in the nonperturbative region
of QCD.

The work has been performed under a partial sponsorship of the Russian Found
of Fundamental Investigations.

\section*{Acknowledgements}
I am thankful to V.A.Novikov for his reading of the manuscript and for the
discussion of some aspects of this work.

I am also indebted to L.B.Okun for his permanent kind interest to my works.

\newpage
\begin{tabular}{lp{12cm}}
{\bf Fig.~1.}& a) Differential crossections of the reaction
$\pi^-p \to \eta n$ at small $q^2_\perp$.
The data \cite{1} V.D.~Apel et al. The solid curve (16).\\
& b) Differential crossections of the reaction
$\pi^-p \to \pi^0 n$ at small $q^2_\perp$.
The data \cite{1} V.D.~Apel et al. The solid curve (16).
\end{tabular}

\begin{tabular}{lp{12cm}}
{\bf Fig.~2.}& Differential crossections $\pi^-p \to K^0\Lambda^0$ -
black circles and $\pi^-p \to K^0\sigma^0$ - white circles.\\
\end{tabular}

\begin{tabular}{lp{12cm}}
{\bf Fig.~3.}& Space-time scheme of the reaction $\pi^-p \to \pi^0n$
in $^3P_0$ model with flux-tube. The arrow $\curvearrowleft$
corresponds to the spin direction along axis +Z.\\
\end{tabular}

\begin{tabular}{lp{12cm}}
{\bf Fig.~4.}& $^3P_0$ transition $\
\bar{u} u \to \bar{d} d$ for the case
of conservation (a),
nonconservation (b) of $l_Z, S_Z$. The straight-line
arrows label 3-momenta of quarks.
\end{tabular}

\begin{tabular}{lp{12cm}}
{\bf Fig.~5.}& The "instanton" transition $\bar{u} u \to s \bar{s}$.
\end{tabular}
\end{document}